\documentstyle[sprocl,epsfig]{article}

\def\Journal#1#2#3#4{{#1} {\bf #2}, #3 (#4)}

\def\NPB{{\em Nucl. Phys.} B}
\def\PLB{{\em Phys. Lett.}  B}
\def\PRL{\em Phys. Rev. Lett.}
\def\PRD{{\em Phys. Rev.} D}

\begin{document}
%
\thispagestyle{empty}
\onecolumn
\begin{flushright}
\large
RIKEN-BNL preprint \\
June 2000
\end{flushright}
\vspace{1.2cm}
 
\renewcommand{\thefootnote}{\fnsymbol{footnote}}
\setcounter{footnote}{1}
\begin{center}
\begin{LARGE}
\begin{bf}
Prompt Photon Production in \\
Polarized Hadron Collisions\footnote{
\begin{large}
Talk presented at the ``8th International Workshop on Deep-Inelastic 
Scattering'' (DIS2000), 25-30 April 2000, Liverpool, UK\\
\end{large}}
 
\end{bf}
\end{LARGE}

\vspace*{1.8cm}
{\Large Werner~Vogelsang\footnote{
\begin{large}
Thanks to RIKEN, Brookhaven National Laboratory and the U.S.
Department of Energy (contract number DE-AC02-98CH10886) for
providing the facilities essential for the completion of this work
\end{large}}}
 
\vspace*{4mm}
 
\begin{large}
{RIKEN-BNL Research Center, Brookhaven National Laboratory}
\vspace*{2mm}
Upton, NY 11973, U.S.A.\\[3pt]
\vspace*{2mm}
E-mail: {\tt wvogelsang@bnl.gov}
\end{large}
\vspace*{2.cm}
 
%
 
{\Large \bf Abstract}
\end{center}
\vskip 3mm
 
\noindent
We consider spin asymmetries for prompt photon 
production in collisions of longitudinally polarized hadrons. This reaction 
will be a key tool at the BNL-RHIC $\vec{p}\vec{p}$ collider for determining 
the gluon spin density in a polarized proton. We study the effects of QCD 
corrections, such as all-order soft-gluon `threshold' resummations.
 
\normalsize
\newpage        

\title{PROMPT PHOTON PRODUCTION IN POLARIZED \\ HADRON COLLISIONS}
\vskip -1mm
\author{W. VOGELSANG}

\address{RIKEN-BNL Research Center, Brookhaven National Laboratory,\\
Upton, NY 11973, U.S.A.}

\maketitle\abstracts{We consider spin asymmetries for prompt photon 
production in collisions of longitudinally polarized hadrons. This reaction 
will be a key tool at the BNL-RHIC $\vec{p}\vec{p}$ collider for determining 
the gluon spin density in a polarized proton. We study the effects of QCD 
corrections, such as all-order soft-gluon `threshold' resummations.}
\vskip -2mm
Prompt-photon production, $pp,p\bar{p},pN\rightarrow \gamma 
X$, has been a classical tool for constraining the unpolarized gluon 
density. At leading order, a photon can be produced in the reactions 
$qg\to\gamma q$ and $q\bar{q}\to\gamma g$, giving rise to a 
distinct clean signal. Using {\em polarized} proton beams at RHIC offers 
a very promising method~\cite{berger89,frixione99,bland99} 
to measure the {\em spin-dependent} gluon density in the nucleon, $\Delta g$, 
which is currently one of the most interesting quantities in nucleon 
structure. 

\begin{figure}[t]
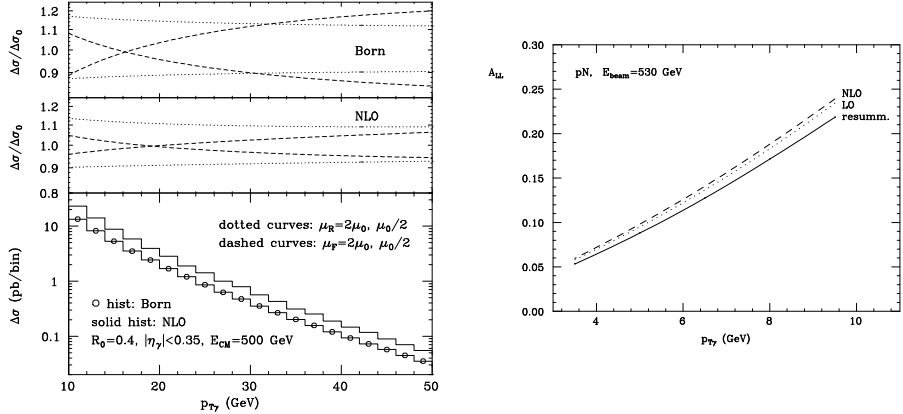

\vskip -0.7cm
\epsfig{figure=liv_p_fig1.eps,height=5.5cm}
\vskip -5cm
\hskip 6.4cm
\epsfig{figure=liv_p_fig2.eps,height=5.5cm,angle=90}
\vskip 0.7cm
\caption{Left: polarized prompt photon cross section in \protect{$\vec{p}
\vec{p}$} collisions at RHIC, at Born and NLO level, as function of
photon transverse momentum. The upper half shows the scale dependence. 
The polarized parton densities were taken from~$\!{}^4$; for further details, 
see~$\!{}^2$. Right: spin asymmetry \protect{$A_{LL}$} at LO, NLO,
and including NLL threshold resummation.
\label{fig:f1}}
\vskip -3mm
\end{figure}

A recent thorough theoretical next-to-leading order (NLO) QCD
study~\cite{frixione99} for prompt photon production at polarized RHIC 
also addressed the unwanted background from photons produced in jet 
fragmentation, when a parton, resulting from a QCD reaction, fragments 
into a photon plus a number of hadrons. In particular, the interplay 
between the fragmentation contribution and the chosen type of `photon 
isolation cut', to be imposed in experiment~\cite{bland99} in order to 
reduce the background from $\pi^0$ decay photons, was studied. By comparing 
to lowest-order results, it was found in~\cite{frixione99} that 
the QCD corrections to the polarized cross section are sizeable and reduce
the dependence of the theory predictions on the factorization and 
renormalization scales; see Fig.~\ref{fig:f1}.

In the unpolarized case, a pattern of disagreement between theoretical
predictions and experimental data for prompt photon production has been 
observed in recent years~\cite{abe94,e706,ua6}. The main problems reside
in the fixed-target region, where NLO theory dramatically underpredicts 
some data sets~\cite{e706,ua6}. At collider energies, as relevant to RHIC, 
there is less reason for concern, but also here the agreement is not 
satisfactory. Various improvements of the theoretical framework have been 
developed. One of them resorts to applying `threshold' resummation 
to the prompt photon cross section~\cite{LOS}, which organizes to all orders 
in $\alpha_s$ large logarithmic corrections to partonic hard scattering, 
associated with emission of soft gluons. It leads to a 
significant, albeit not sufficient, enhancement of the theory prediction in 
the fixed-target regime at large values of $p_T/\sqrt{s}$, accompanied by a 
dramatic reduction of scale dependence~\cite{CMNOV}. Thanks to the universal 
structure of soft-gluon emission, it is straightforward to apply threshold 
resummation to the polarized cross section. Fig.~\ref{fig:f1} shows the 
resulting effects on the {\em spin asymmetry} $A_{LL}$, for a `toy' example 
that assumes a fictitious polarized set-up of the E706 
experiment~\cite{e706}. Details are as in~\cite{CMNOV}. Even 
though Fig.~\ref{fig:f1} does not directly 
refer to the case of RHIC, it is good news that resummation effects 
cancel to a large extent in $A_{LL}$ for our present example.
 
\section*{Acknowledgments}
I am grateful to S.~Frixione and G.~Sterman for fruitful collaborations.

\end{document}